\documentclass[3p]{elsarticle}
\usepackage{times}
\usepackage{epsfig}
\usepackage{setspace}
\usepackage{array}
\usepackage{multirow}
\usepackage{booktabs}
\usepackage{color}
\usepackage{amsfonts}
\usepackage{threeparttable}
\usepackage{amsmath,bm}
\usepackage{lscape}
\usepackage{algorithm}
\usepackage{algpseudocode}
\usepackage{siunitx}
\usepackage{hyperref}
\newcommand{\tabincell}[2]{\begin{tabular}{@{}#1@{}}#2\end{tabular}}
\journal{Journal of \LaTeX\ Templates}

\biboptions{authoryear}

\begin{document}

\begin{frontmatter}

  \title{Domain adaptation based self-correction model for COVID-19 infection segmentation in CT images}

  \author[mymainaddress,mythirdaddress]{Qiangguo~Jin}

  \author[mysecondaryaddress]{Hui~Cui}

  \author[mythirdaddress]{Changming~Sun}

  \author[mymainaddress,myfourthaddress]{Zhaopeng~Meng}

  \author[myfifthaddress]{Leyi~Wei}
  
  \author[mymainaddress]{Ran Su}

  \address[mymainaddress]{School of Computer Software, College of Intelligence and Computing, Tianjin University, Tianjin, China}
  \address[mysecondaryaddress]{Department of Computer Science and Information Technology, La Trobe University, Melbourne, Australia}
  \address[mythirdaddress]{CSIRO Data61, Sydney, Australia}
  \address[myfourthaddress]{Tianjin University of Traditional Chinese Medicine, Tianjin, China}
  \address[myfifthaddress]{School of Software, Shandong University, Shandong, China}

  \begin{abstract}
    The capability of generalization to unseen domains is crucial for deep learning models when considering real-world scenarios.
    However, current available medical image datasets, such as those for COVID-19 CT images, have large variations of infections and domain shift problems.
    To address this issue, we propose a prior knowledge driven domain adaptation and a dual-domain enhanced self-correction learning scheme. Based on the novel learning schemes, a domain adaptation based self-correction model (DASC-Net) is proposed for COVID-19 infection segmentation on CT images. DASC-Net consists of a novel attention and feature domain enhanced domain adaptation model (AFD-DA) to solve the domain shifts and a self-correction learning process to refine segmentation results. The innovations in AFD-DA include an image-level activation feature extractor with attention to lung abnormalities and a multi-level discrimination module for hierarchical feature domain alignment.
    The proposed self-correction learning process adaptively aggregates the learned model and corresponding pseudo labels for the propagation of aligned source and target domain information to alleviate the overfitting to noises caused by pseudo labels. Extensive experiments over three publicly available COVID-19 CT datasets demonstrate that DASC-Net consistently outperforms state-of-the-art segmentation, domain shift, and coronavirus infection segmentation methods. Ablation analysis further shows the effectiveness of the major components in our model. The DASC-Net enriches the theory of domain adaptation and self-correction learning in medical imaging and can be generalized to multi-site COVID-19 infection segmentation on CT images for clinical deployment.
  \end{abstract}


\end{frontmatter}

\section{Introduction}
\label{sec:introduction}
Deep learning (DL) methods achieved remarkable success with the fundamental hypothesis that the training data and test data are from an identical distribution~\citep{robinson2020image}. However, in real-world scenarios, this hypothesis is often violated in the fields of natural image and medical image processing. A typical situation in the medical field is the usage of various image datasets, which can differ significantly in data distribution due to different hospitals, scanner vendors, imaging protocols, patient populations, etc~\citep{abramoff2018pivotal,dou2018pnp}. As the distribution of domain changes, a well-trained system may fail to produce precise predictions for unseen data with domain shift.

To tackle this issue, domain adaptation (DA) algorithms normally learn to align source and target data in a domain-invariant discriminative feature space~\citep{dou2018pnp,chen2019crdoco,zhang2018fully,saporta2020esl,chen2020unsupervised,zhu2019boundary}. 
Many of those DA methods embed the source and target data into the latent space to obtain similar distributions. By such embedding, the well-trained model on the source domain can be applied to the target domain~\citep{wang2019domain}.
However, obstacles exist in those types of DA methods. Firstly, most of the DA models focus on narrowing the distribution variations in higher-order latent space while neglecting the influence of hierarchical low-level semantic features in antecedent. Secondly, few DA models integrate prior knowledge in the feature extraction and discrimination process. Prior knowledge, however, can provide supervision especially when there is a limited amount of data available for training. Finally, for those DA based segmentation models, the segmentation masks, or the so-called pseudo labels, are obtained based on DA results of the target domain. The source and aligned target domains, however, are not revisited for potential refinement over the pseudo labels.

\begin{figure}
  \centering
  \includegraphics[scale=0.45]{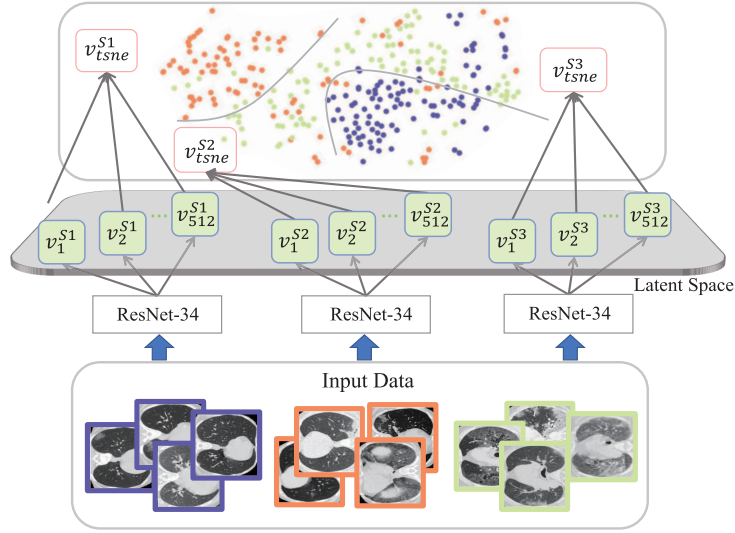}
  \caption{Illustration of the domain shift problem in the three datasets of COVID-19 used in this work. Latent space features extracted by ResNet-34 are visualized by t-SNE, where each cluster has 100 randomly selected samples from each domain. $v^{S1}_{i}$ denotes the $i$-th feature vector extracted from domain $S1$ in the latent space, $S1$, $S2$, and $S3$ denote different domains, and the feature vector $v^{S1}_{tsne}$ is embedded by t-SNE.}
  \label{fig:overview of domain shift problem}
\end{figure}

To address those issues, we propose a novel domain adaptation based self-correction learning (DASC) algorithm. In practice, the novel algorithm is applied to the coronavirus infection segmentation task.
Coronavirus (COVID-19) infection 
segmentation from computed tomography (CT) images is an emerging issue with the outbreak of COVID-19~\citep{zhu2020novel}. The automatic segmentation of COVID-19 infections on CT images, however, is challenging because of several reasons.
Firstly, the data volume of COVID-19 CT images is limited. Besides, datasets with pixel-level annotations are even harder to obtain when the world, especially experts with domain knowledge, is fighting the COVID-19 pandemic.
Secondly, domain shifts exist in the currently available public COVID-19 CT image datasets as illustrated in Fig.~\ref{fig:overview of domain shift problem}.
Thirdly, large variations exhibit in COVID-19 infections on CT images, such as irregular shapes, indistinct boundaries, and inhomogeneous intensity distributions~\citep{wang2020noise}. It makes the automated localization and segmentation of abnormalities even challenging.

\begin{figure}
  \centering
  \includegraphics[scale=0.45]{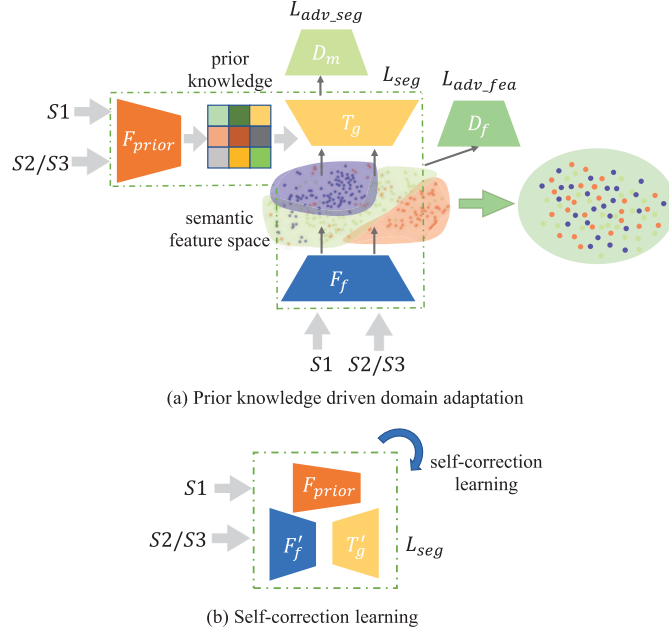}
  \caption{An overview of the proposed (a) prior knowledge-driven domain adaptation and (b) dual-domain enhanced self-correction learning scheme to address the domain shift problem in semantic feature space, especially when there are limited amounts of data for training. $F_{f}$ and $T_{g}$ denote the feature extractor and decoder. $D_{m}$ and $D_{f}$ are the mask-level and hierarchical feature level discriminators. $F_{prior}$ is prior knowledge extractor. $F^{'}_{f}$ and $T^{'}_{g}$ refer to the adaptively refined $F_{f}$ and $T_{g}$. $L_{seg}$, $L_{adv\_seg}$, and $L_{adv\_fea}$ denote the segmentation loss, mask domain adversarial loss, and feature domain adversarial loss function.}
  \label{fig:overview of proposed motivation}
  \end{figure}

To make full use of the limited amount of available datasets and address the domain shift problems, we propose a novel prior knowledge driven domain adaptation and a dual-domain enhanced self-correction learning scheme (Fig.~\ref{fig:overview of proposed motivation}). Given data from both source ($S1$) and target ($S2$/$S3$) domains, we firstly extract features under the supervision of prior knowledge from image-level labels. To solve the domain shift problem in semantic feature space, we introduce hierarchical feature-level alignment and discrimination. Then the initial segmentation, i.e., pseudo labels, is obtained after alignment. Finally, we propose a dual-domain enhanced self-correction learning mechanism to adaptively refine the pseudo labels by learning from both source and target domain samples. In practice, we propose a domain adaption based self-correction model (DASC-Net). The unique contributions of the proposed algorithm are summarized below.

\begin{itemize}
  \item Theoretically, we propose a novel prior knowledge driven domain adaptation and a dual-domain enhanced self-correction learning scheme. The proposed DA scheme is capable of minimizing the impact of domain shift in the mask-level domain and hierarchical feature-level domain. In the meanwhile, the potential refinement over the pseudo labels is formulated by the self-correction learning scheme.
  \item Practically, we propose a novel attention and feature domain enhanced DA model (AFD-DA) to improve the segmentation performance, especially when there are variations of infections and domain shifts. The innovations include a class activation map (CAM)~\citep{zhou2016learning} based segmentation to emphasize lung abnormalities, and hierarchical feature-level alignment and discrimination.
  \item To fully utilize the limited amount of available data for training, we propose a dual-domain enhanced self-correction learning algorithm to refine the segmentation results. The self-correction learning algorithm enables model optimization and pseudo label aggregation during training cycles. Besides, the source and target domains are co-learnt in self-correction to minimize the misleading supervision from noises caused by pseudo labels.
  \item Extensive experiments using three public COVID-19 CT datasets and comprehensive evaluations demonstrated improved segmentation results over state-of-the-art methods for medical image segmentation and domain adaptation. Besides, the extensive experiments show that our proposed DASC-Net fills the gap between theoretical and practical implications. The source code will be released publicly at GitHub~\footnote{https://github.com/qgking/DASC\_COVID19.git}.
\end{itemize}

In the following sections, we present related work in Section~\ref{sec:related_works}. The methodologies are given in Section~\ref{sec:methodology}, where we first introduce the basic DA model, followed by the novel AFD-DA model, self-correction learning, and implementation details. Datasets and pre-processing strategy are given in Section~\ref{sec:data_set}. Section~\ref{sec:exp_and_res} presents and discusses experimental results with a conclusion in Section~\ref{sec:conclusion}.

\section{Related works}
\label{sec:related_works}
Recent methods for semantic segmentation are mainly based on deep learning~\citep{lv2020cross,zhou2020automatic,qiu2020miniseg,shi2020review,jin99cascade,zlocha2019improving,dou2020unpaired}, and some of which have achieved state-of-the-art performances in COVID-19 analysis and domain adaptation fields.

\subsection{Computer-aided COVID-19 analysis from CT images}
Recent applications of AI technologies to COVID-19 CT images fall into three categories: severity assessment, COVID-19 screening from other types of pneumonia, and infection segmentation~\citep{shi2020review}.

\textbf{Severity assessment}:
\citet{he2020synergistic} proposed a multi-instance based architecture to predict the severity of COVID-19. The synergistic learning framework achieved joint segmentation of lung lobes and classification of COVID-19. 
\citet{tang2020severity} trained a random forest model using severity-related features to automatically quantify the severity.

\textbf{COVID-19 screening}:
\citet{shi2020large} developed an infection size aware random forest model where location-specific features were extracted and sorted for classification.
Attention-based 3D convolutional neural network (CNN) was developed by~\citet{ouyang2020dual} to improve the screening performance with the focus on COVID-19 infected regions.
\citet{kang2020diagnosis} proposed a multi-view representation learning model to extract complementary types of features for COVID-19 diagnosis.

\begin{figure*}
  \centering
  \includegraphics[scale=0.50]{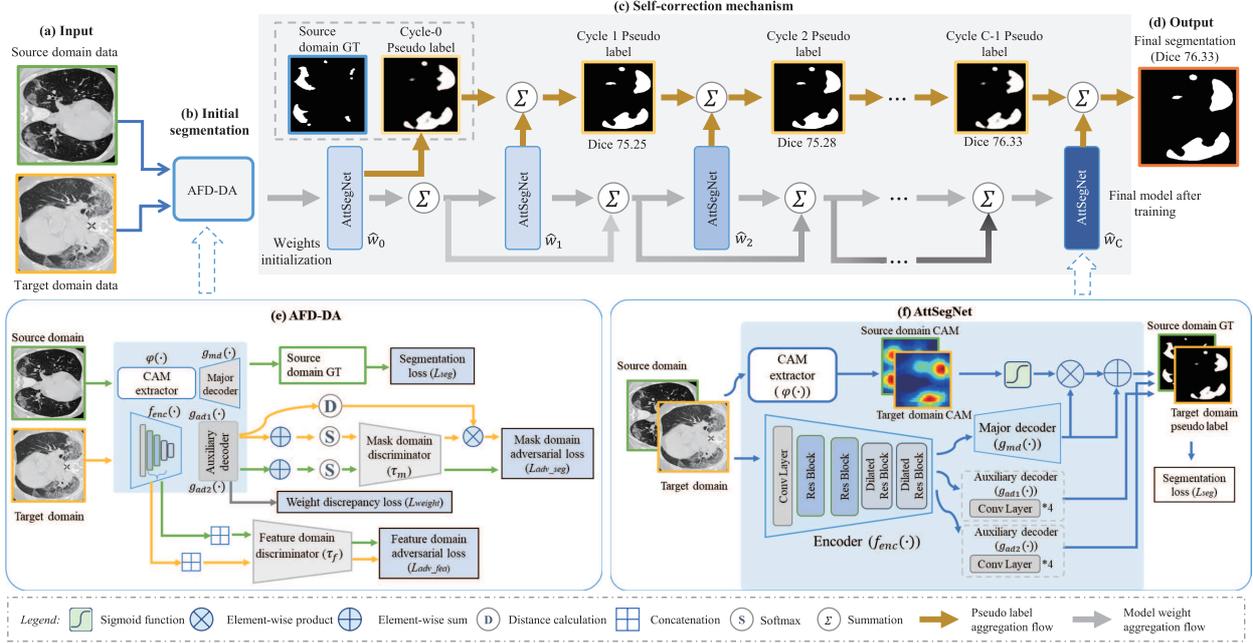}
  \caption{Overview of the proposed DASC-Net for COVID-19 CT segmentation. Given (a) input data, (b) AFD-DA is firstly trained to solve the domain shift problem and obtain initial segmentation results. The detailed structure of ADF-DA is in (e). (c) The self-correction learning process iteratively aggregates (f) AttSegNet model and pseudo labels until the final cycle to generate (d) final segmentation results. AttSegNet is initialized by the generator of ADF-DA.}
  \label{fig:overview of architecture}
\end{figure*}

\textbf{Infection segmentation}:
To automatically segment COVID-19 infected regions, \citet{zhou2020automatic} developed a DL model using aggregated residual transformations and attention mechanism.
\citet{chen2020residual} extracted contextual features by incorporating spatial and channel attentions to a U-Net architecture. The proposed method was evaluated on a COVID-19 CT segmentation dataset with 473 CT slices.
\citet{zhou2020rapid} proposed a machine-agnostic method that can segment and quantify the infection regions on multi-sites CT scans.
A joint classification and segmentation (JCS) system for real-time and interpretable COVID-19 diagnosis was developed by~\citet{wu2020jcs}.
Because public COVID-19 dataset is limited and hard to obtain, \citet{qiu2020miniseg} developed a lightweight deep learning model for efficient COVID-19 segmentation while avoiding overfitting.
\citet{fan2020inf} proposed a semi-supervised learning model to tackle the challenge of a limited amount of data for training.

Given a small amount of data for training which may also exhibit domain shifts, the machine learning (ML) methods could have overfitting problems, especially for data-hungry DL models. Hence, how to make the best use of a limited amount of cross-site data for COVID-19 analysis is an emerging and challenging problem to be solved.

\subsection{DA based semantic segmentation}
Most of the ML models for semantic segmentation are based on the hypothesis that the training and test data have an identical distribution~\citep{lv2020cross}. The hypothesis, however, is not always true in real-world data. When transferring knowledge from a label-rich source domain to a label-scarce target domain, it is common to find the discrepancy from the training to the test stage. The DA based technique aims to rectify this discrepancy and tune the models towards better generalization at the testing step~\citep{patel2015visual}. When the data distribution gap between source and target domains is narrowed, an improved generality can be obtained.

Over the past few years, DA techniques attract intensive interests in semantic segmentation~\citep{lv2020cross,luo2019significance,tsai2019domain,zhu2017unpaired,Chen_2020_CVPR,li2019bidirectional,li2019self,du2019ssf,zheng2019unsupervised}. Generally, DA models can be divided into two categories~\citep{lv2020cross}: domain-invariant representation learning and pseudo label guided learning.

\subsubsection{Domain-invariant representations learning}
One of the key motivations behind this category of DA based semantic segmentation is to learn domain invariant representations. Most of the recent studies are based on domain adversarial learning and generative adversarial networks (GANs). For instance,
\citet{luo2019significance} combined an information bottlenecked strategy and an adversarial learning framework for feature-space DA semantic segmentation.
\citet{tsai2019domain} proposed an adversarial adaptation scheme, which makes the feature representations of target patches closer to the distributions of source patches in clustered spaces.

CycleGAN~\citep{zhu2017unpaired} provides an alternative solution for DA by translating the source images into target-style images.
For instance, \citet{Chen_2020_CVPR} proposed a domain adaptive image-to-image translation architecture for out-of-domain samples.
\citet{li2019bidirectional} proposed a bidirectional learning framework for self-supervised DA segmentation.
Unlike domain adversarial learning based methods, \citet{lv2020cross} demonstrated that pivot information could improve the performance of DA based segmentation, which served as a bridge to share common knowledge between source and target domains.

\subsubsection{Pseudo label guided learning}
The pseudo-labelling method, which is a typical technique in semi-supervised learning~\citep{lee2013pseudo}, can also be used to solve domain shifts. 
The pseudo label normally refers to the label for target data~\citep{li2019self}. For instance, 
\citet{zheng2019unsupervised} proposed a model to mine the target domain knowledge and fine-tune unlabelled target data in a self-training approach. 
\citet{du2019ssf} introduced a progressive confidence strategy taking full advantage of pseudo labels via adversarial learning without global feature alignment.
In~\citep{saporta2020esl}, entropy is used as a confidence indicator to improve the quality of pseudo labels in an entropy-guided self-supervised learning (ESL) model.

A recent self-correction training strategy~\citep{li2019self} is proposed to improve the segmentation performance during different training cycles by self-supervision. 
The apparent drawback of using this approach is that if the pseudo labels of target domain contain noises, the noisy information may mislead the correction of pseudo labels and result in overfitting to the noise. However, our approach leverages both source and target domain information for improved self-correction, especially when there is a limited amount of data with domain shifts.


\section{Methods}
\label{sec:methodology}
The network architecture of DASC-Net is given in Fig.~\ref{fig:overview of architecture}. 
The DASC-Net consists of an attention and feature domain enhanced DA model (AFD-DA) for initial segmentation and a self-correction procedure for segmentation correction and improvement. The AFD-DA has three innovations: a class activation map (CAM)~\citep{zhou2016learning} (i.e., prior knowledge) based attentive segmentation branch to drive the network's attention to lung infections, a discrimination module for hierarchical feature-level domain alignment and discrimination, and a hybrid loss function for optimization. The major components in the self-correction learning process include learning cycles where each cycle has an attention enhanced segmentation network (AttSegNet) and a self-correction mechanism to propagate model parameters and pseudo labels.

\subsection{Baseline DA model via adversarial learning}
\subsubsection{Problem formulation}
Given an image $I_{s} \in \mathbb{R}^{H \times W}$ from infected source domain $I_{S}$ and an image $I_{t} \in \mathbb{R}^{H \times W}$ from target domain $I_{T}$, where $H$ and $W$ are the height and width of the images, $S$ and $T$ are the index set of images in source and target domains respectively, and we use the labelled source domain $\mathcal{D}_{S}=\left\{\left(I_{s}, Y_{s}\right)\right\}_{s \in S}$ and the unlabelled target domain $\mathcal{D}_{T}=\left\{I_{t}\right\}_{t \in T}$ to train a segmentation network, where $Y_{s} \in \left\{  0,1\right \} ^{H \times W}$ denotes the corresponding label of $I_{s}$. The $I_{s}$ is first passed to the segmentor with $Y_{s} \in \left\{  0,1\right \} ^{H \times W}$. Afterwards, the segmentor infers the segmentation output $P_{t}$ for image $I_{t}$. The goal of domain adaptation is to achieve the desirable semantic segmentation performance on the target domain.
\subsubsection{Basic DA framework}
\label{sec:basicDA}
We exploit the DA model in~\citep{luo2019taking} as the basic network for a better local semantic consistency enforcement during the procedure of global alignment, which is composed of a generator $G$ and a mask domain discriminator $\tau_{m}(\cdot)$. The generator $G$ is divided into feature encoder $f_{enc}(\cdot)$ and two decoders, termed as $g_{ad1}(\cdot)$ and $g_{ad2}(\cdot)$. In our experiments, the dilated ResNet-34~\citep{he2016deep,yu2017dilated} is adopted as the feature encoder. During the co-training procedure~\citep{zhou2005tri}, the weights of $g_{ad1}$ and $g_{ad2}$ are diversified by a cosine distance cost function.

High-level feature maps are extracted from $f_{enc}(\cdot)$ after feeding $I_{s}$ from the source domain. Those high-level feature maps serve as the input of the two decoders, i.e, $g_{ad1}$ and $g_{ad2}$, to yield the final predictions $P^{1}_{s}$ and $P^{2}_{s}$. $P^{1}_{s}$ and $P^{2}_{s}$ are used to penalize the network for learning a better prediction under the supervision of $Y_{s}$. The segmentation loss function $L_{seg}$ is defined as:
\begin{equation}
  L_{seg}=\mathbb{E}[L_{ce}(P^{1}_{s},Y_{s})+L_{ce}(P^{2}_{s},Y_{s})],
  \label{eq:loss_seg}
\end{equation}
where $L_{ce}$ denotes the cross-entropy loss, and $P^{1}_{s}$ and $P^{2}_{s}$ are obtained based on $g_{ad1}(f_{enc}(I_{s}))$ and $g_{ad2}(f_{enc}(I_{s}))$ respectively. In the meantime, $P^{1}_{s}$ and $P^{2}_{s}$ are also treated as mask domain inputs to the discriminator $\tau_{m}$ for mask domain adversarial learning.

On the other hand, $P^{1}_{t}$ and $P^{2}_{t}$ can also be obtained by $G$. Apart from mask domain adversarial loss, pixel-wise discrepancy $Dis(P^{1}_{t},P^{2}_{t})$ between $P^{1}_{t}$ and $P^{2}_{t}$ is measured by a cosine similarity as:
\begin{equation}
  Dis(P^{1}_{t},P^{2}_{t})= 1-{\rm cos}(P^{1}_{t},P^{2}_{t}).
  \label{eq:cosine_dis}
\end{equation}
In summary, the adversarial learning procedure can be formulated as:
\begin{equation}
  \begin{array}{l}
    L_{adv\_seg}=\mathbb{E}\left[\log \left(\tau_{m}\left(P^{1}_{s}+P^{2}_{s}\right)\right)\right]+ \\
    \mathbb{E}\left[\left(\lambda_{dis} Dis(P^{1}_{t},P^{2}_{t})\right) \log \left(1-\tau_{m}\left(P^{1}_{t}+P^{2}_{t}\right)\right)\right],
  \end{array}
  \label{eq:loss_adv_seg}
\end{equation}
where $\lambda_{dis}$ is a weight factor.
Each pixel on the segmentation map is differently weighted with respect to the mask domain adversarial loss~\citep{luo2019taking}.

The basic DA model also provides the weight discrepancy loss on $g_{ad1}$ and $g_{ad2}$ for providing two different views on features~\citep{luo2019taking}. The $L_{weight}$ is formulated as:
\begin{equation}
  L_{weight}=\frac{\overrightarrow{w_{1}} \cdot \overrightarrow{w_{2}}}{\|\overrightarrow{w_{1}}\|\|\overrightarrow{w_{2}}\|},
  \label{eq:weight_loss}
\end{equation}
where $\overrightarrow{w_{1}}$ and $\overrightarrow{w_{2}}$ are obtained by flattening and concatenating the weights of the convolutional layers of $g_{ad1}$ and $g_{ad2}$.

The final training objective $L_{BASE}$ in the base DA model can be summaried as:
\begin{equation}
  L_{BASE}= L_{seg}+\lambda_{weight} L_{weight}+ \lambda_{adv\_seg} L_{adv\_seg},
  \label{eq:loss_taining}
\end{equation}
where $\lambda_{weight}$ and $\lambda_{adv\_seg}$ are weight factors.

\subsection{Prior knowledge driven domain adaptation}
As there are large variations between the COVID-19 source domain and target domain, the baseline DA model may fail to identify infected lung regions which are tiny or have high similarities with healthy tissues. Besides, the basic DA only penalizes mask-domain alignment. However, there still exist domain misalignment between features in antecedent. Thus, we propose an AFD-DA model to improve the performance of the basic DA.

We hypothesize that a DA model with attention to lung abnormities and hierarchical feature domain alignment and discrimination can enhance the segmentation performance. Fig.~\ref{fig:overview of architecture}(e) gives the architecture of AFD-DA. Compared with the basic DA, the AFD-DA includes an attentive segmentation branch with a CAM extractor and a major decoder, and a feature-level domain alignment and discrimination module where there are feature encoder, two auxiliary decoders, and a mask-level discriminator.

\subsubsection{Prior knowledge driven segmentation branch}
\begin{figure}
  \centering
  \includegraphics[scale=0.6]{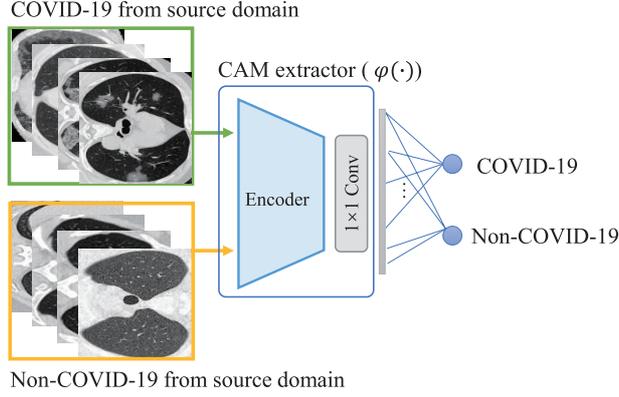}
  \caption{CAM extractor for attention to lung abnormalities. The classification network is pre-trained using COVID-19/Non-COVID-19 data from the source domain.}
  \label{fig:overview of cam feature extractor}
\end{figure}

In the segmentation branch, we firstly design a CAM enhanced encoder with attention to lung abnormalities. This is achieved by pre-training a classification network using image-level labels as shown in Fig.~\ref{fig:overview of cam feature extractor}. Since CAM can highlight image regions associated with image-level labels, we use positive samples (CT slices with infections, i.e., $I_{S}$) and negative samples (CT slices without infections) from the source domain as the training dataset. The classification network consists of a dilated ResNet-34 encoder and a $1 \times 1$ convolutional layer. The pre-trained network is referred to as CAM extractor $\varphi$.

As CAM may contain non-infected regions which can mislead the segmentation~\citep{xie2020mutual}, we propose an online CAM attention based decoder to improve the conventional decoders $g_{ad1}$ and $g_{ad2}$ in the basic DA. Inspired by residual learning~\citep{he2016deep}, the idea is that if soft CAM can be constructed as the attention map, the performance shall be at least better than a decoder without attention. 

Given DeepLab V3+~\citep{chen2017deeplab} as the major segmentor $g_{md}$, we modify output $P^{0}_{s}$ of major segmentor $g_{md}$ as 
\begin{equation}
  P_{s}^{0}=\left(1+\sigma\left(Up\left(\varphi\left(I_{s}\right)\right)\right)\right) g_{md}\left( f_{enc}\left(I_{s}\right) \right),
  \label{eq:online_cam_att_seg}
\end{equation}
where $\sigma(\cdot)$ denotes the sigmoid function, $Up(\cdot)$ is a bilinear upsample function to interpolate the CAM to the input resolution, $\sigma\left(Up\left(\varphi\left(I_{s}\right)\right)\right)$ is within the range of [0, 1]. After combining Equation~(\ref{eq:loss_seg}) and Equation~(\ref{eq:online_cam_att_seg}), we can rewrite the final segmentation loss function as:
\begin{equation}
  L_{seg}=\mathbb{E}[ \lambda_{seg} L_{ce}(P^{0}_{s},Y_{s})+L_{ce}(P^{1}_{s},Y_{s})+L_{ce}(P^{2}_{s},Y_{s})],
  \label{eq:loss_seg_final}
\end{equation}
where $\lambda_{seg}$ is a weight factor to magnify the impact of the major decoder, the two decoders are denoted as auxiliary decoders.

\subsubsection{Hierarchical feature-level alignment of semantic feature}
According to~\citep{luo2019taking}, lower $Dis(P^{1}_{t},P^{2}_{t})$ values indicate less severe domain shifts and larger overlap between source and target domain distributions. It seems that the domain shift problem is solved in the latent space. However, due to the hierarchical architecture of the convolutional layers, the features extracted from a specific layer heavily rely on the activations from its previous layers and vice versa~\citep{dou2018pnp}. Thus, if a model only focuses on the high-level latent space alignment, domain shift problems may still exist during the lower-level feature extractions. The problem may be magnified when the misaligned low-level features are propagated to the deeper layers for final integration.

To overcome this problem, and inspired by deep supervision~\citep{lee2015deeply}, we propose hierarchical feature-level domain alignment discrimination. 
As shown in Fig.~\ref{fig:overview of architecture}(e), there are a feature extractor $f_{enc}$ for high-level and low-level knowledge encoding and feature-level discriminator $\tau_{f}$ to facilitate the conventional mask-level discriminator. We aggregate and concatenate the outputs, which are the input of the feature-level discriminator $\tau_{f}$, from every residual block in $f_{enc}$ except the last one. Given $f_{enc}$, the feature-level domain alignment can be supervised by the gradients from $\tau_{f}$ to $f_{enc}$ as
\begin{equation}
  \begin{split}
    L_{adv\_fea}=&\mathbb{E}\left[\log \left(\tau_{f}\left(f_{enc}^{1}(I_{s}) ,f_{enc}^{2}(I_{s}),f_{enc}^{3}(I_{s})\right)\right)\right] \\
    + &\mathbb{E}\left[\log \left(1-\tau_{f}\left(f_{enc}^{1}(I_{t}) ,f_{enc}^{2}(I_{t}) ,f_{enc}^{3}(I_{t})\right)\right)\right],
  \end{split}
  \label{eq:loss_adv_feature}
\end{equation}
where $f_{enc}^{i}(I_{s})$ denotes the output of the $i$-th residual block in feature extractor $f_{enc}$.

\subsection{Pseudo label refinement via self-correction learning}
To fully utilize the available information for training and decrease the impact of noise in the self-correction process, we propose a dual-domain enhanced self-correction learning mechanism. Because the AFD-DA model addressed the domain shift problem, our motivation is that the label-rich source domain can be blended with the target domain to correct the noise caused by pseudo labels. As shown in Fig.~\ref{fig:overview of architecture}, the process consists of $C$ training cycles where each cycle has an AttSegNet. The self-correction mechanism aggregates the model and propagates the updated labels.

\begin{algorithm}[h]
  \caption{Self-correction learning mechanism}
  \label{code:self_learning}
  \begin{algorithmic}[1]
    \Require
    $\hat{\omega}_{0}$: initialized model weight by $f_{enc}$, $g_{md}$, $g_{ad1}$, $g_{ad2}$, and $\varphi$ after domain adaptation;
    $\hat{y}_{0}^{a}$: augmented initial pseudo label generated by $\hat{\omega}_{0}$;
    $Ep$: epoch in each cycle;
    $C$: total number of cycles;
    \Ensure
    Optimal $\hat{\omega}_{C}$ \\
    Construct a training dataset $\mathcal{D}_{training}$ using $\mathcal{D}_{S}=\left(I_{S}, Y_{S}\right)$ and $\mathcal{D}_{T}=\left(I_{T}, \hat{Y}_{0}^{a}\right)$
    \For{each $c\in [1,C]$}
    \For{each $e\in [1,Ep]$}
    \State Update the learning rate;
    \State Calculate loss $L_{seg}$ by Equation~(\ref{eq:loss_seg_final});
    \EndFor
    \State Update weight $\hat{\omega}_{c+1}$ for initializing the weight of the next cycle $c+1$ by Equation~(\ref{eq:model_aggregation});
    \State Generate pseudo label $\hat{Y}_{c+1}$ using $\hat{\omega}_{c}$;
    \State Augment pseudo label $\hat{Y}_{c+1}^{a}$ using $\hat{\omega}_{c}$;
    \State Pseudo label refinement by Equation~(\ref{eq:pseudo_label_aggregation});
    \State Re-construct a training dataset $\mathcal{D}_{training}$ using $\mathcal{D}_{S}=\left(I_{S}, Y_{S}\right)$ and $\mathcal{D}_{T}=\left(I_{T}, \hat{Y}_{c+1}^{a}\right)$
    \EndFor

  \end{algorithmic}
\end{algorithm}

\subsubsection{Self-correction mechanism}
Given $C$ training cycles and a set of optimal segmentation models $S=\left\{\hat{\omega}_{0}, \hat{\omega}_{1}, \ldots, \hat{\omega}_{C}\right\}$ where each of them is obtained after a training cycle, $\hat{\omega}_{0}$ denotes the well-trained generator in our AFD-DA model. At the beginning of the $c$-th cycle, we initialize model $\hat{\omega}_{c}$ by aggregating model $\hat{\omega}_{c-1}$ with $\hat{\omega}_{0}$. The aggregation is formulated as:
\begin{equation}
  \hat{\omega}_{c}=\frac{c}{c+1} \hat{\omega}_{c-1}+\frac{1}{c+1} \hat{\omega}_{0},
  \label{eq:model_aggregation}
\end{equation}
where $c$ is in $\{1, \ldots, C\}$ and $C$ is set as 9 in our experiments.
The pseudo-code of the self-learning algorithm is given in Algorithm~\ref{code:self_learning}. During the self-correction integration, the pseudo labels are progressively improved in terms of performance and generability.

\subsubsection{Pseudo label aggregation and AttSegNet}
As shown in Fig.~\ref{fig:overview of architecture}(f), AttSegNet consists of a CAM extractor $\varphi$, feature extractor $f_{enc}$, CAM enhanced decoder $g_{md}$, and two auxiliary decoders $g_{ad1}$ and $g_{ad2}$, which are structurally identical to the generator in AFD-DA. In each training cycle, the segmentation model takes mixed source data with ground-truth and target data with pseudo labels as input. In practice, we update the pseudo label of the target domain at the end of each cycle. To further alleviate the incorrect segmentation in pseudo labels from the target domain, we augment the target domain by horizontal and vertical flipping, and then aggregate those augmented predictions to increase the reliability of pseudo labels.

With $Y^{a}=\left\{\hat{Y}_{0}^{a}, \hat{Y}_{1}^{a}, \ldots, \hat{Y}_{C}^{a}\right\}$ denoting the augmented set of all the pseudo labels after each cycle, pseudo labels are updated as follows based on our model aggregation strategy:
\begin{equation}
  \hat{Y}_{c}^{a}=\frac{c}{c+1} \hat{Y}_{c-1}^{a}+\frac{1}{c+1} \hat{Y}_{0}^{a},
  \label{eq:pseudo_label_aggregation}
\end{equation}
where $\hat{Y}_{0}^{a}$ is generated by the initial model $\hat{\omega}_{0}$.

\subsection{Network architecture and loss function}
In our DASC-Net, there are three networks requiring optimization: a CAM extractor for online CAM generation, AFD-DA for domain adaptation training, and AttSegNet for pseudo label and model aggregation in self-correction.
\subsubsection{Network architecture}
Even though performance can be improved with a deeper depth of network architecture, a deeper network may hamper the training process and cause problems of vanishing and exploding gradients~\citep{he2016deep}. To address this issue, \citet{he2016deep} proposed a deep residual learning scheme, where the deep residual unit makes the deep network easy to train and alleviates the problem of vanishing gradients~\citep{he2016deep}. Because of these advantages, we use ResNet as the backbone encoders in our models. Since there are limited amount of available COVID-19 data and computational resources, we choose the 34-layer ResNet. Besides, atrous convolutions with different dilation rates are employed within each residual block to increase the receptive field of each layer for better and multi scale feature representation.
The encoders in all three models in our work are identical to the dilated ResNet-34~\citep{he2016deep,yu2017dilated} and initialized with the weight pre-trained on ImageNet~\citep{deng2009imagenet}.

The mask domain discriminator $\tau_{m}$ consists of 5 convolutional layers with $4 \times 4$ kernels, channel numbers of $\{ 64,128,256,512,1\}$, and stride of 2. Each convolutional layer is followed by a Leaky-ReLU with a negative slope of 0.2 except the last layer. The last layer is an up-sampling layer to restore the output to the size of the input domain. The structure of feature domain discriminator $\tau_{f}$ is the same as $\tau_{m}$ but without the last up-sampling layer, and the channel numbers are set as $\{256,128,64,64,1\}$.
\subsubsection{Loss function}
We use the cross-entropy loss as the objective function to train the CAM extractor:
\begin{equation}
  \mathrm{CE}=-\frac{1}{N} \sum_{i=1}^{N} \sum_{m=1}^{M} p_{i}^{m} \log \hat{p}_{i}^{m},
  \label{eq:cam_generator_loss}
\end{equation}
where $\hat{p}_{i}^{m}$ denotes the probability of an image sample $i$ belonging to class $m$, $p_{i}^{m}$ denotes the ground-truth class, $N$ indicates the total number of samples, and $M$ represents the total class type of a slice, i.e., $M=2$.

In adversarial learning, the $G$ of AFD-DA is trained against an adversary with the two discriminators $\tau_{m}$ and $\tau_{f}$. The training objective of AFD-DA is formulated as:
\begin{equation}
  \begin{split}
    L_{da}= & L_{seg}+\lambda_{weight} L_{weight}+ \\
    & \lambda_{adv\_seg} L_{adv\_seg} + \lambda_{adv\_fea} L_{adv\_fea},
  \end{split}
  \label{eq:loss_taining_adv_final}
\end{equation}
where $\lambda_{weight}, \lambda_{adv\_seg}$, and $\lambda_{adv\_fea}$ are weight factors, $L_{seg}$ is calculated by Equation~(\ref{eq:loss_seg_final}), $L_{weight}$ is defined by Equation~(\ref{eq:weight_loss}), $L_{adv\_seg}$ and $L_{adv\_fea}$ are formulated in Equation~(\ref{eq:loss_adv_seg}) and Equation~(\ref{eq:loss_adv_feature}), respectively.

AttSegNet is initialized by the generator of AFD-DA. We penalize the network using Equation~(\ref{eq:loss_seg_final}) with source and target domain data.

\section{Dataset and preprocessing}
\label{sec:data_set}
\subsection{Dataset}
We collect COVID-19 CT datasets from three sites to evaluate the performance of DASC-Net.

\subsubsection{Coronacases Initiative and Radiopaedia (CIR)}
CIR released 20 public 3D CT scans without manual segmentations. \citet{ma2020towards} provided manually segmented lung and infections by radiologists. Among the images, 10 scans are $630 \times 630$ in-plane resolution, while the other 10 scans are of $512 \times 512$. In total, there are 3,520 axial CT slices, including 1,844 slices with annotated COVID-19 infections (considered as positive 2D samples) and 1,676 slices without annotations (considered as negative 2D samples).

\subsubsection{Italian Society of Medical and Interventional Radiology (ISMIR)}
ISMIR provided 100 axial CT images from 60 patients with COVID-19~\footnote{http://medicalsegmentation.com/covid19/}. The manual segmentation of COVID-19 infected areas, including ground-glass opacity (GGO), consolidation, and pleural effusion, was performed by a radiologist. The image size is $512 \times 512$ pixels and has been greyscaled and compiled into a single NIFTI file.

\subsubsection{MOSMEDDATA}
Research and Practical Clinical Center for Diagnostics and Telemedicine Technologies of the Moscow Health Care Department~\citep{morozov2020mosmeddata} provided 50 CT volumetric scans and corresponding annotations by experts. For each scan, the annotations of GGO and consolidation are recorded by non-zero values in a binary mask. In total, there are 2,049 axial CT slices, including 785 COVID-19 positive and 1,264 negative samples. It is noted that the proportion of the samples containing GGO and consolidation is less than 25\%. The in-plane resolution of all the images is $512 \times 512$.

In the experiments, we use the CIR data as the source domain, termed as COVID-19-S. The ISMIR and MOSMEDDATA data are used as target domains which are denoted by COVID-19-T1 and COVID-19-T2 respectively.

\subsection{Data pre-processing and augmentation}
Pre-processing includes image normalization and patch generation. All the CT images were normalized by windowing operation of [-1,250, 250] Hounsfield Units (HU) values, and then linearly scaled to [0, 1] by min-max normalization. Secondly, we cropped the images by localizing lung regions. Manual-segmented lungs were provided in COVID-19-S and COVID-19-T1. To obtain the lung mask in COVID-19-T2, we trained a 3D U-Net~\citep{ronneberger2015u} using COVID-19-S. Finally, the image patches were obtained by cropping the original CT image using a bounding box outside the lung. 

Data augmentation transformations included random flippings in vertical and horizontal directions, affine transformations using translation within [0.01, 0.01], scaling within [0.8, 1.2], shearing within [-10, 10], and rotating by up to $90^\circ$. Finally, we down-sampled the infected 2D CT slices and infection masks into $320 \times 320$ pixels for efficient computation.


\section{Experimental results and discussion}
\label{sec:exp_and_res}
\subsection{Implementation and parameter settings}
The DASC-Net model was implemented using the PyTorch~\citep{paszke2017automatic} library and trained on an NVIDIA 1080Ti GPU. For the loss function, the hyper-parameters $\lambda_{seg}$, $\lambda_{weight}$, $\lambda_{adv\_seg}$, $ \lambda_{adv\_fea}$, and $\lambda_{dis}$ are set to 3, 0.01, 0.001, 0.001, and 10 respectively.
Adam optimizer, with a batch size of 4, is applied to minimize the objectives in DASC-Net. The learning rate for the discriminators is set to $\num{1e-4}$, while the others are set as $\num{2.5e-4}$. All the optimizers are decayed by a polynomial learning rate policy, where the initial learning rate is multiplied by $\left(1-\frac{iter}{total\_{iter}}\right)^{power}$ with $power$ at 0.9. The total number of training iterations is set to $100* (iter\ per\ epoch)$, i.e., 100 epoch, for AFD-DA. For AttSegNet, the total number of cycles $C$ is set to 9 and the total number of epochs in each cycle is set to 2. The training of the CAM extractor stopped after one epoch obtains a map which can totally cover the infections.

\subsection{Comparison methods and evaluation metrics}
The segmentation models are validated using a comprehensive list of evaluation metrics including sensitivity (SEN), specificity (SPC), Jaccard (JA), Dice coefficient (Dice), and Hausdorff distance (HD)~\citep{HausdorffDistance}.

To demonstrate the effectiveness of our DASC-Net, we compare with several segmentation models, which can be categorized as U-Net based model for medical image segmentation, DA based model for semantic segmentation, and most recent segmentation models published for COVID-19 infection segmentation.

\textbf{U-Net based model}: The U-Net based models are most widely used in medical image analysis. As there are few published peer-reviewed methods for COVID-19 infection segmentation, we firstly compared with U-Net based state-of-the-art models. U-Net~\citep{ronneberger2015u} is one of the most widely used CNN models for medical image segmentation. The architecture captures contextual features and utilizes a symmetric expanding path for precise localization.
Compared to U-Net, U2-Net~\citep{qin2020u2}\footnote{https://github.com/NathanUA/U-2-Net} is able to capture more contextual information from different scales with the receptive fields of different sizes in the Residual U-blocks. The residual U-blocks increase the depth of the network with few computational cost and shows promising results for segmentation.
U-Net++~\citep{zhou2018unet++}\footnote{https://github.com/4uiiurz1/pytorch-nested-unet} is a nested architecture, which is designed for medical image segmentation, and has been evaluated across multiple tasks.
Taking advantage of full-scale skip connections and deep supervisions, the U-Net 3+~\citep{huang2020unet}\footnote{https://github.com/ZJUGiveLab/UNet-Version} is proposed, which shows desirable results on liver segmentation.

\textbf{DA based model}:
We also compare with recent DA based models including AdaptSegNet~\citep{tsai2018learning}, ADVENT~\citep{vu2019advent}. Different from our AFD-DA model, those models lack strict hierarchical feature alignment and discrimination. AdaptSegNet~\citep{tsai2018learning}\footnote{https://github.com/wasidennis/AdaptSegNet} is an adversarial learning method for domain adaptation in the context of semantic segmentation. A multi-level adversarial algorithm is proposed to perform output space domain adaptation.
Entropy of the pixel-wise predictions is fully utilized by ADVENT~\citep{vu2019advent}\footnote{https://github.com/valeoai/ADVENT}, which provides an alternative solution for DA.

\textbf{COVID-19 infection segmentation model}:
Semi-Inf-Net~\citep{fan2020inf} is compared as it is a most recent segmentation model proposed for COVID-19 infection segmentation. The model uses a semi-supervised approach to address the problem of training on a limited amount of data. The model was only validated on COVID-19-T1.

\subsection{Comparison results}

\begin{table}[]
  \centering
  \begin{threeparttable}
    \caption{Results on the COVID-19-T1 dataset for DASC-Net and SOTA methods. The best results are shown in bold font.}
    \label{table:comparative results of DASC on COVID-19-T1}
    \renewcommand\arraystretch{1.1}
    \renewcommand\tabcolsep{12.5pt}
    \begin{tabular}{ccccccccc}
      \toprule
      \textbf{Method}                              & \textbf{Dice~(\%)} & \textbf{SEN~(\%)} & \textbf{SPC~(\%)} &  \textbf{HD} & \textbf{JA~(\%)}  \\ \hline
      U-Net~\citep{ronneberger2015u}   & 70.30  & 71.83  & 98.27   & 80.80  & 55.72    \\
      U2-Net~\citep{qin2020u2}         & 69.58  & 77.16  & 97.26   & 87.82  & 55.03     \\
      U-Net++~\citep{zhou2018unet++}   & 67.46  & 75.67  & 96.73   & 89.63  & 52.69       \\
      U-Net 3+~\citep{huang2020unet}   & 69.60  & 78.00  & 97.09   & 91.52  & 54.84      \\
      AdaptSegNet\tnote{*} ~\citep{tsai2018learning} & 71.29  & 76.62   & 97.64  & 83.53  & 56.83    \\
      ADVENT\tnote{*} ~\citep{vu2019advent}          & 72.01  & 78.39   & 97.66  & 77.49  & 57.82  \\  \hline
      Ours (AFD-DA)\tnote{*}                           & 74.33  & 80.92   & 97.72   & 71.41  & 60.52     \\
      Ours (DASC-Net)\tnote{*}      & \textbf{76.33}    & \textbf{83.24}   & \textbf{97.82}  & \textbf{65.31}    & \textbf{62.97} 
      \\
      \bottomrule
    \end{tabular}
    \begin{tablenotes}
      \footnotesize
      \item[*] DA based method
    \end{tablenotes}
  \end{threeparttable}
\end{table}

\begin{table}[]
  \centering
  \begin{threeparttable}
    \caption{Results on the COVID-19-T2 dataset for DASC-Net and SOTA methods. The best results are shown in bold font.}
    \label{table:comparative results of DASC on COVID-19-T2}
    \renewcommand\arraystretch{1.1}
    \renewcommand\tabcolsep{12.5pt}
    \begin{tabular}{ccccccccc}
      \toprule
      \textbf{Method}                              & \textbf{Dice~(\%)} & \textbf{SEN~(\%)} & \textbf{SPC~(\%)} &  \textbf{HD} & \textbf{JA~(\%)} \\ \hline
      U-Net~\citep{ronneberger2015u}   & 51.95  & 61.99 & 99.76  & 95.09  & 39.56    \\
      U2-Net~\citep{qin2020u2}         & 56.91  & 66.52 & 99.76  & 71.55  &  44.12   \\
      U-Net++~\citep{zhou2018unet++}   & 54.00  & 73.88 & 99.58  & 95.72   &  40.41    \\
      U-Net 3+~\citep{huang2020unet}   & 55.46   & 73.80 & 99.61 & 78.58   &  42.18    \\
      AdaptSegNet\tnote{*} ~\citep{tsai2018learning} & 56.18  & 69.87   & 99.70    & \textbf{64.97}    & 43.15        \\
      ADVENT\tnote{*} ~\citep{vu2019advent}          & 56.40     & 69.00     & 99.70      & 66.78   & 43.36    \\ \hline
      Ours (AFD-DA)\tnote{*}                           & 59.04  & \textbf{75.17}   & 99.74   & 74.14  & 45.42   \\
      Ours (DASC-Net)\tnote{*}      & \textbf{60.66}    & 72.44   & \textbf{99.78}     & 75.62    & \textbf{46.96}  
      \\
      \bottomrule
    \end{tabular}
    \begin{tablenotes}
      \footnotesize
      \item[*] DA based method
    \end{tablenotes}
  \end{threeparttable}
\end{table}

\begin{table}[]
  \centering
  \begin{threeparttable}
    \caption{Results on the 50 slices from COVID-19-T1 dataset for DASC-Net and Semi-Inf-Net. The best results are shown in bold font.}
    \label{table:comparative with inf-net}
    \renewcommand\arraystretch{1.1}
    \renewcommand\tabcolsep{5.5pt}
    \begin{tabular}{ccccc}
      \toprule
      \textbf{Method}    & \textbf{Dice~(\%)} & \textbf{SEN~(\%)} & \textbf{SPC~(\%)}  \\ \hline
      Semi-Inf-Net~\citep{fan2020inf}   & 76.4  & 79.7  & 96.3      \\
      Ours      & \textbf{77.0}    & \textbf{81.2}   & \textbf{98.0} 
      \\
      \bottomrule
    \end{tabular}
  \end{threeparttable}
\end{table}

\subsubsection{Quantitative results}
The evaluation results over COVID-19-T1 using our DASC-Net, DASC-Net without self-correction, referred to as AFD-DA, and all the other models are given in Table~\ref{table:comparative results of DASC on COVID-19-T1}. As shown in the table, our AFD-DA outperformed all the other methods for all the evaluation metrics especially HD, Dice, and SEN. The results demonstrated the effectiveness of AFD-DA in addressing domain shifts. The self-correction mechanism further boosted the performance of AFD-DA by 2\% and 6.1 in terms of Dice and HD, and 2.32\%, 0.1\%, and 2.45\% with respect to SEN, SPC, and JA.

The comparison results over COVID-19-T2 are summarized in Table~\ref{table:comparative results of DASC on COVID-19-T2}. As shown in this table, the DASC-Net achieved the best results, followed by AFD-DA. We also found that the overall experimental results by all the models on COVID-19-T2 were worse than those on COVID-19-T1. This may be due to the fact that the infections have large variations between COVID-19-T2 and COVID-19-S. For instance, in the COVID-19-T2 samples, image regions with GGO and pulmonary parenchymal typically cover less than 25\% of the whole lung. In the source domain, however, there are not too many CT slices with less than 25\% infected areas. Besides, our empirical finding is that the less the portion of the infections, the more influence on the segmentation results. Nevertheless, our model outperformed all the methods in comparison over the cases with smaller infection regions.

When comparing with Semi-Inf-Net~\citep{fan2020inf}, we found that the experimental results in~\citep{fan2020inf} were using 50 samples from COVID-19-T1 only. Thus, we selected the same 50 slices as reported in~\citep{fan2020inf}~\footnote{https://github.com/DengPingFan/Inf\-Net} and re-calculated the segmentation results for a fair comparison. The comparison results are given in Table~\ref{table:comparative with inf-net}. 

\begin{figure}
  \centering
  \includegraphics[scale=0.91]{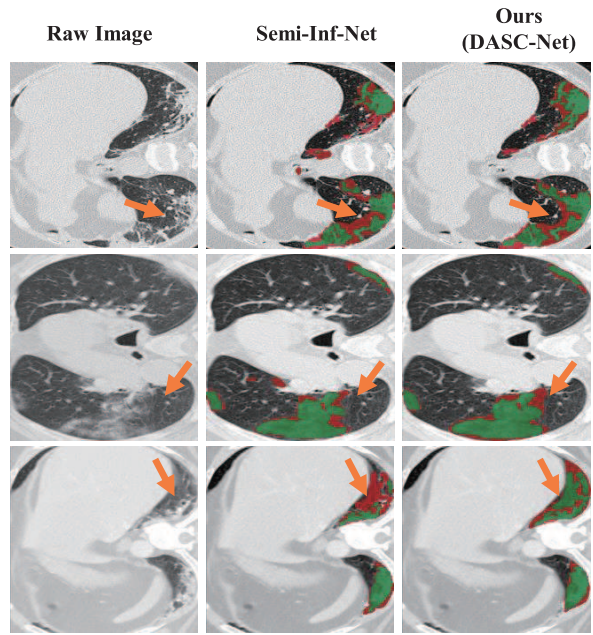}
  \caption{Three examples of COVID-19 infection segmentation against Semi-Inf-Net. The segmentation results from Semi-Inf-Net were downloaded from the authors' GitHub repository. The false predictions, i.e., false-positive and false-negative, are shown in red while the correct predictions are in green. The significant improvement by our model is marked with orange arrows.}
  \label{fig:overview of comparison with semi-inf-net}
\end{figure}

Our DASC-Net achieved a Dice of 77.0\%, a SEN of 81.2\%, and a SPC of 98.0\% which outperformed the published results of Semi-Inf-Net by 0.6\%, 1.5\%, and 1.7\% respectively. As shown in Fig.~\ref{fig:overview of comparison with semi-inf-net}, the proportion with red color indicates false predictions. It reveals that the DASC-Net provides a better segmentation with higher confidence.
It is also noted that the Semi-Inf-Net was trained using the rest 50 slices from COVID-19-T1. Our DASC-Net, however, was trained using source domain data without fine-tuning on the 50 slices from the target domain. Compared with Semi-Inf-Net, our model is more applicable to real-world scenarios, especially when the training and testing datasets are from different sites.

Our primary finding is that our DASC-Net substantially alleviates the domain shift problem and achieves superior performance in segmenting COVID-19 infection when compared to the U-Net based models (i.e., U-Net~\citep{ronneberger2015u}, U2-Net~\citep{qin2020u2}, U-Net++~\citep{zhou2018unet++}, and U-Net 3+~\citep{huang2020unet}). Secondly, the DA based models (i.e., AdaptSegNet~\citep{tsai2018learning} and ADVENT~\citep{vu2019advent}) gain better quantitative results than the U-Net based models. Finally, our DASC-Net outperforms the COVID-19 infection segmentation model (i.e., Semi-Inf-Net~\citep{fan2020inf}) without fine-tuning on extra data, which is extremely important at the early stage of COVID-19 outbreak when limited well-annotated data samples can be acquired.

\begin{figure*}
  \centering
  \includegraphics[scale=0.6]{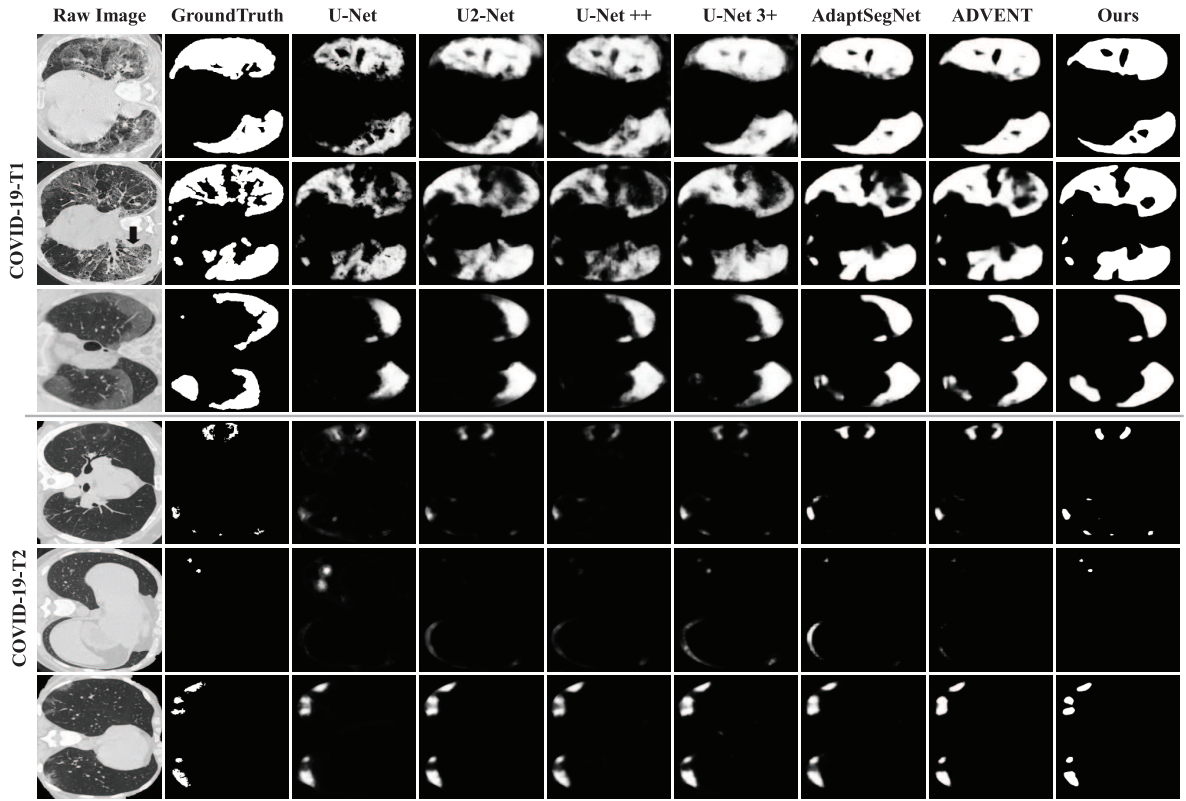}
  \caption{Visual comparison of COVID-19 infection segmentation against other methods on two target datasets.}
  \label{fig:overview of comparison}
\end{figure*}

\subsubsection{Qualitative results}
Six segmentation results on COVID-19-T1 and COVID-19-T2 are given in Fig.~\ref{fig:overview of comparison}. The segmentation results of DASC-Net are the closest to manual segmentations whether the regions to be segmented are large or small, as shown by the top three and bottom three rows. It is also found that the DA based models, including AdaptSegNet~\citep{tsai2018learning} and ADVENT~\citep{vu2019advent}, outperformed those U-Net based models without domain adaptation.

\begin{table*}[]
  \centering
  \begin{threeparttable}
    \caption{Ablation studies of our DASC-Net on the COVID-19-T1 dataset. A component marked by \boldsymbol{$\surd$} means that the model contains this component. The best results are shown in blod font.}
    \label{table:ablation of DASC}
    \renewcommand\arraystretch{1.1}
    \renewcommand\tabcolsep{3.0pt}
    \begin{tabular}{ccccccccc}
      \toprule
      \textbf{\tabincell{c}{Basic                                                                                                                                \\ DA}} & \textbf{\tabincell{c}{CAM attention\\ enhanced}} & \textbf{\tabincell{c}{Feature-level  \\ alignment}} & {\textbf{Self-correction}} & \textbf{Dice~(\%)} & \textbf{SEN~(\%)}   & \textbf{SPC~(\%)}  & \textbf{HD}    & \textbf{JA~(\%)}  \\ \hline
      \boldsymbol{$\surd$} &      &     &     & 73.65 & 79.97 & 97.76 & 69.70 & 59.64 
      \\
      \boldsymbol{$\surd$} & \boldsymbol{$\surd$} &        &     &  74.36    &  80.02    & 97.87  &  68.46  &   60.51    \\
      \boldsymbol{$\surd$} &   & \boldsymbol{$\surd$} &    & 73.84
                            & 80.96
                            & 97.61
                            & 68.99                & 59.79
      \\
      \boldsymbol{$\surd$} & \boldsymbol{$\surd$} & \boldsymbol{$\surd$} &                      & 74.33
                            & 80.92
                            & 97.72
                            & 71.41                & 60.52
      \\
      \boldsymbol{$\surd$} & \boldsymbol{$\surd$} & \boldsymbol{$\surd$} & \boldsymbol{$\surd$} & \textbf{76.33}
                            & \textbf{83.24}
                            & \textbf{97.82}
                            & \textbf{65.31}       & \textbf{62.97}
      \\
      \bottomrule
    \end{tabular}
  \end{threeparttable}
\end{table*}

\subsection{Ablation studies for effectiveness validation}
Ablation study was performed to evaluate the contributions of major components in DASC-Net. We first obtain the performance of the baseline DA model~\citep{luo2019taking}. Afterwards, we gradually added the newly proposed components to the baseline model. The experimental results on COVID-19-T1 are presented in Table~\ref{table:ablation of DASC}.

\subsubsection{Contributions of prior knowledge driven segmentation branch and hierarchical feature-level alignment of semantic features in AFD-DA}
The baseline DA model~\citep{luo2019taking} achieved a Dice of 73.65\%. When an online CAM based attentive segmentation branch was added to the baseline, the experimental results were improved. The newly proposed feature-level domain alignment also contributed to the experimental results where the Dice value further increased to 73.84\%. The SEN and HD metrics were improved by 0.99\% and 0.71 respectively. When considering both CAM based attentive segmentation branch and feature domain alignment module, the performance was further improved to a Dice of 74.33\%.

The experimental results verified our hypothesis that although domain shift problems are tackled in a high-level latent space by the basic DA, the problem still exists in the low-level feature space. The proposed feature-level domain alignment module enabled our model to resolve the feature-level domain shifts, as demonstrated by the improved segmentation results.

\begin{figure*}
  \centering
  \includegraphics[scale=0.48]{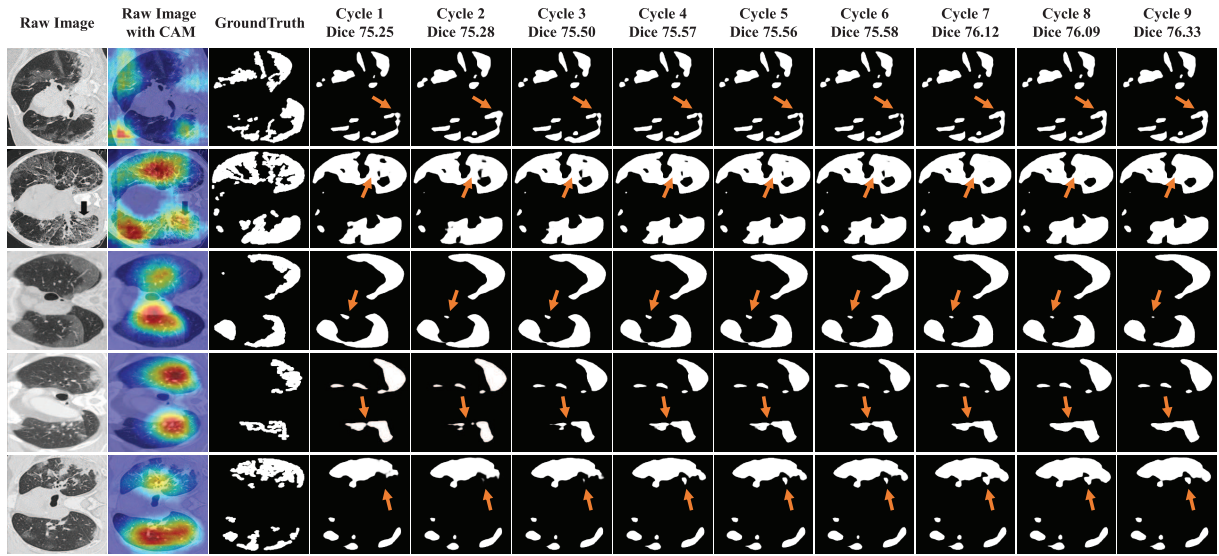}
  \caption{Illustration of segmentation map in 9 cycles on COVID-19-T1. The refinement is marked with orange arrows. }
  \label{fig:overview of 9 cycles output}
\end{figure*}

\subsubsection{Effectiveness of self-correction learning}
When examining the contributions of the self-correction algorithm, Table~\ref{table:ablation of DASC} shows that the most noticeable improvement was in HD, followed by JA, Dice, and SEN.

We also investigate the segmentation performance in different cycles during the self-correction learning process. The detailed segmentation results and the corresponding Dice scores in each cycle are given in Fig.~\ref{fig:overview of 9 cycles output}. As shown in this figure, with the aggregation of the model and pseudo labels, the segmentation results gradually increased from a Dice of 75.25\% in Cycle 1 to a Dice of 76.33\% in the final cycle. More importantly, the segmentation results on small or isolated infected regions were improved as indicated by the orange arrows in Fig.~\ref{fig:overview of 9 cycles output}.

\section{Conclusion}
\label{sec:conclusion}
We propose a novel domain adaptation based self-correction learning (DASC-Net) model. The DASC-Net utilizes the prior domain knowledge and enhances the distribution similarity of the semantic features by hierarchical feature-level discrimination, and adaptively refines the pseudo labels by self-correction learning mechanism. In practice, CAM drives the segmentation network to emphasize lung abnormalities. The new hierarchical feature-level discriminator complements mask-level discrimination, which enhances feature domain alignment. The self-correction learning mechanism alleviates the misleading supervision caused by noises in pseudo labels. Extensive experiments on multi-sites public COVID-19 datasets demonstrate that our model outperforms the state-of-the-art methods. Our model and framework can be generally applied to other tasks when there are domain shift problems and lack of sufficient dataset for training, which bridges the theory-practice gap. The limitation of our method is that we assume that all of the source data samples are fully annotated. However, in the early stage of the COVID-19 outbreak, the amount of well-annotated data samples is still limited, and the performance of DA methods can degrade substantially with fewer labeled samples. Our future work includes the exploration of DASC-Net for few-shot learning tasks.

\section*{Acknowledgment}
This work is supported by the National Natural Science Foundation of China [Grant No. 61702361, 62072329 and 62071278], the Science and Technology Program of Tianjin, China [Grant No. 16ZXHLGX00170], the Natural Science Foundation of Tianjin [Grant No. 18JCQNJC00800 and 18JCQNJC00500], the National Key Technology R\&D Program of China [Grant No. 2018YFB1701700], and the program of China Scholarships Council.

\bibliography{elsarticle-template}

\end{document}